\documentclass[12pt]{amsart}
\usepackage[mathscr]{eucal}
\usepackage{amsmath,amsfonts}

\parskip=\smallskipamount
                                                                                
\newtheorem{prop}{Proposition}
\newtheorem{thm}{Theorem}

\newtheorem{lem}{Lemma}
\newtheorem{algo}{Algorithm}
\newtheorem{defi}{Definition}

\makeatletter
\@addtoreset{equation}{section}
\makeatother

\newcommand{\bC}{{\mathbf C}}

\newcommand{\bN}{{\bf N}}

\newcommand{\cB}{{\mathcal B}}

\newcommand{\cH}{{\mathcal H}}
\newcommand{\cK}{{\mathcal K}}
\newcommand{\cL}{{\mathcal L}}

\newcommand{\cS}{{\mathcal S}}

\newcommand{\la}{\langle \;}
\newcommand{\ra}{ \; \rangle}

\newcommand{\pf}{\noindent {\it Proof:} }

\newcommand{\noi}{\noindent}

\newdimen\expt
\def\boxit#1{\setbox0\hbox{$\displaystyle{#1}$}
      \hbox{\lower.4\expt
 \hbox{\lower3\expt\hbox{\lower\dp0
      \hbox{\vbox{\hrule height.4\expt
 \hbox{\vrule width.4\expt\hskip3\expt
      \vbox{\vskip3\expt\box0\vskip2\expt}%
 \hskip3\expt\vrule width.4\expt}\hrule height.4\expt}}}}}}
\begin{document}
\pagestyle{plain}

\bigskip

\title {Contractions, Matrix Paramatrizations, and Quantum Information}

\author{M. C. Tseng\\
Department of Mathematical Sciences\\
University of Texas at Dallas}

\begin{abstract}
In this note, we discuss dilation-theoretic matrix parametrizations of contractions and positive matrices. These parametrizations are then applied
to some problems in quantum information theory. First we establish some properties of positive maps, or entanglement witnesses.
Two further applications, concerning concrete dilations of completely positive maps, in particular quantum operations, are given. 
\end{abstract} 

\maketitle

\section{Introduction}

It is well-known that positive operator-matrices, or more generally positive kernels, can be parametrized by contractions \cite{Co}. In this
paper, we show that analogous results can be obtained for matrix contractions. A common feature of these parametrizations is that, while 
the explicit expressions may seem intricate, due to their combinatorial nature they can be easily understood by inspecting the so-called 
{\sl lattice diagrams}. These diagrams will be used repeatedly to illustrate accompanying results.

The organization of this note is as follows. The structure of row and column contractions are discussed first. They already possess an elegant
combinatorial structure and play a central role in our parametrizations. Next we consider matrix contractions. 
The $2 \times 2$ matrix contractions were already characterized in \cite{Gheon}. Here we extend the description to matrices of arbitrary size and 
point out the combinatorial aspect of this parametrization. Then the special case of unitary matrices is examined. We also review the 
parametrization of positive matrices. While the definitive treatment of the positive case is \cite{Co}, our discussion differs slightly
in some minor technical details. Turning to applications, we first obtain some properties of positive maps. The structure of positive maps and
contractions is applied to show that general positive maps are more than merely positive when restricted to certain subsets of positive matrices.
Results of this type  were obtained in \cite{Choi} and the parametrization of positive matrices allows one to explore their extensions in a non-ad hoc way.
By the correspondence between positive maps and entanglement witnesses \cite{Horodecki}, we thus show that certain families of bipartite mixed states
are separable. The last two applications concerns the unitary dilation of completely positive maps on matrix algebras. While a celebrated result by 
Stinespring \cite{Stinespring} shows that such dilations always exist, the paramatrization of contractions allows one to give a concrete constructive 
procedure for such dilations.

The results on positive maps are first stated in operator-theoretic terms before being placed in physical context. The last two applications are phrased
more directly in the language of quantum information theory. For general background in quantum information, we refer the reader to \cite{Werner} and
\cite{nie}.

Although we only consider matrices of finite size, all parametrizations described in this paper can be extended to (semi-)infinite matrices,
where convergence is given by an appropriate operator topology.

\section{Row and Column Contractions}

The facts outlined below can be found in \cite{Co}, in which the structure of such contractions plays a crucial
role in obtaining a parametrization for positive kernels, the {\sl Schur-Constantinescu parameters}. Our presentation here is
different in that the notion of {\sl defect spaces} is dispensed with. In \cite{Co}, the defect spaces
are used in forcing the uniqueness of certain operator angles. Rather, we identify the unique positive square roots explicitly via partial
isometries.\\

In the following, $\cH$ and $\cH_i$ denote Hilbert spaces, and $\cL( \cH_1 , \cH_2)$ the bounded operators from $\cH_1$ to $\cH_2$.

\begin{lem}
Let $X \in \cL( \cH_1 , \cH_2)$ and $Y \in \cL( \cH_1 , \cH_2)$ be bounded operators between Hilbert spaces. $X^* X \leq Y^*Y$ iff there exist a 
contraction $\Gamma : \cH_2 \rightarrow \cH_2$ such that $\Gamma X =Y$.
\end{lem}

\pf Suppose $X^* X \leq Y^* Y$. Define $\Gamma' : $ by $\Gamma' X h = Y h$ 
on $Ran X$, the range of X. Extending by continuity to the closure of $Ran X$ and then by 0 to the orthogonal compliment gives a contraction $\Gamma$. 
The expression $X^* X \leq Y^*Y$ implies $Ker Y \subset Ker X$. So $\Gamma$ is a contraction satisfying $\Gamma X =Y$. The converse is also 
straightforward. $\Box$\\

%The action of Y on the orthogonal compliment of ran(X) needs to be considered here.

\noindent This leads to the following well-known fact regarding the freedom in the square-roots of a bounded operator. 

\begin{lem}
If $X^* X = Y^*Y$, then there exist a partial isometry $V$ such that $V X = Y$. Equivalently, $X^* V^* = Y^*$.
\end{lem}

\pf It is clear that we can take $V$ to be the $\Gamma$ defined above.  $\Box$\\

The partial isometry $V$ is unique if the condition $Ker V \subset (Ran L)^{\perp}$ is imposed. This uniqueness condition will be assumed throughout. 
The operator $V$ is unitary if both $Ran X$ and $Ran Y$ are dense. When $\cH_1 = \cH_2 = \cH_3 = \cH$ is finite dimensional, $V$ can aslo be assumed unitary.
Given a positve operator $A \in \cL(\cH)$ with unique positive square root $A^{ \frac{1}{2} }$, every $L$ such that $L^* L = A$ is related to 
$A^{ \frac{1}{2} }$ by $A^{ \frac{1}{2} } = V L$, or $A^{ \frac{1}{2} } = L^* V^*$.

These two lemmas will be used to exhibit the structure of row contractions. Before doing so, we first introduce a bit of terminology. For a contraction 
$T \in \cL( \cH_1, \cH_2)$, the positive operator $(I - T^*T)^{\frac{1}{2}} \in  \cL(\cH_1)$ is called the {\sl defect operator} of $T$ and denoted
by $D_{T}$. Similarly, $D_{T^*} = (I - TT^*)^{\frac{1}{2}} \in  \cL(\cH_2)$ is the defect operator of $T^*$. When $\cH_1 = \cH_2$, 
by the continuous functional calculus,

\[
T(I-T^*T) = (I - T T^*)T
\]

\noindent implies $T D_T = D_{T^*} T$. This relation shows that the operator

\[
J(T) =
\left[
\begin{array}{cc}
T & D_{T^*}\\
D_T & -T^*
\end{array}
\right]
\] 

\noindent , called the {\sl Julia operator}  of the contraction $T$, is unitary. The Julia operator can be viewed as the building blocks of
our parametrizations. It is represented by figure 1.\\

\begin{figure}[h]
\setlength{\unitlength}{3000sp}%
\begingroup\makeatletter\ifx\SetFigFont\undefined%

\gdef\SetFigFont#1#2#3#4#5{%
  \reset@font\fontsize{#1}{#2pt}%
  \fontfamily{#3}\fontseries{#4}\fontshape{#5}%
  \selectfont}%
\fi\endgroup%

\begin{picture}(6174,2949)(289,-2323)

%Box 1

{ \put(2500,-200){\vector( 1,-1){1200}}
}%
{ \put(2500,-1400){\vector( 1, 1){1200}}
}%
{ \put(2400,-1600){\framebox(1650,1650){}}
}%
{ \put(2200,-200){\line( 1, 0){2000}}
}%
{ \put(2200,-1400){\line( 1, 0){2000}}
}%
{ \put(2200,-200){\vector( 1, 0){1700}}
}%
{ \put(2200,-1400){\vector( 1, 0){1700}}
}%

{ \put(3100,-170){$T$}
}%

{ \put(2950,-1630){$-T^*$}
}%

{ \put(2370,-490){$D_T$}
}%

{ \put(2370,-1200){$D_{T^*}$}
}%

\end{picture}

\caption{The Julia operator $J(T)$}
\end{figure}
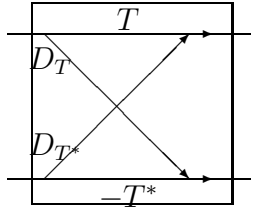

Next we describe row contractions of length two.

\begin{prop} Let $T = [T_1 \; T_2] \in \cL( \cH_1 \oplus \cH_2, \cH)$, then $\|T\| \leq 1$ iff there exist contractions $\Gamma_1 \in
\cL( \cH_1, \cH)$ and $\Gamma_2 \in \cL( \cH_2, \cH)$ such that $T = [\Gamma_1 \; D_{\Gamma_1 ^*} \Gamma_2]$.
\end{prop}

\pf $(\Rightarrow)$ We can take $\Gamma_1$ to be $T_1$. $\|T\| \leq 1$ implies 

\[
I - T T^* = I - \Gamma_1 \Gamma_1^* - T_2 T_2 ^* \geq 0
\] 

\noindent i.e. $D_{\Gamma_1 ^*} ^2 \geq T_2 T_2^*$. By lemma, $ \Lambda D_{\Gamma_1 ^*} = T_2 ^*$. choosing
$\Gamma_2 = \Lambda ^*$ finishes the argument.

$(\Leftarrow)$ Direct computation. $\Box$\\

The defect operators for the contraction $T = [T_1 \; T_2]$ can be directly calculated: 

\[
D_T ^2 = 
\left[
\begin{array}{cc}
D_{\Gamma_1} & 0\\
- \Gamma_2^* \Gamma_1 & D_{\Gamma_1}
\end{array}
\right]
\left[
\begin{array}{cc}
D_{\Gamma_1} & - \Gamma_1^* \Gamma_2\\
0 & D_{\Gamma_1}
\end{array}
\right]
\]

\noindent Invoking lemma, we have 

\[
D_T =
\left[
\begin{array}{cc}
D_{\Gamma_1} & 0\\
- \Gamma_2^* \Gamma_1 & D_{\Gamma_1}
\end{array}
\right]
V
\]

\noindent for some unique partial isometry $V$. Similarly,

\[
D_{T^*}^2 = ( D_{\Gamma_1 ^*} D_{\Gamma_2 ^*} D_{\Gamma_2 ^*} D_{\Gamma_1 ^*} ).
\]

The general structure of row contractions is described by:

\begin{thm}
Let $T = [T_1 \; T_2 \; \cdots ] \in \cL( \oplus \cH_i, \cH)$, then $\|T\| \leq 1$ iff there exist contractions $\Gamma_i \in
\cL( \cH_i, \cH)$ such that 
\[
T = [\Gamma_1, D_{\Gamma_1 ^*} \Gamma_2, D_{\Gamma_1 ^*} D_{\Gamma_2 ^*} \Gamma_3, \cdots, D_{\Gamma_1 ^*} \cdots D_{\Gamma_{n-1} ^*} \Gamma_n].
\]

\noindent Furthermore, the defect operators are of the form $D_T^2 =$

\[
\left[
\begin{array}{cccc}
D_{\Gamma_1}                                                       & 0             & \cdots & 0 \\
- \Gamma_2 ^* \Gamma_1                                             & D_{\Gamma_2}  & \cdots &  0         \\
\vdots                                                             & \vdots        & \ddots & \vdots\\
-\Gamma_n^* D_{\Gamma_{n-1} ^*} \cdots D_{\Gamma_2 ^*} \Gamma_1    & -\Gamma_n^* D_{\Gamma_{n-1} ^*} \cdots D_{\Gamma_{2} ^*} \Gamma_2 & \cdots & D_{\Gamma_n} \\
\end{array}
\right]
\left[
\begin{array}{cccc}
D_{\Gamma_1} & - \Gamma_1 ^* \Gamma_2 & \cdots & -\Gamma_1^* D_{\Gamma_2 ^*} \cdots D_{\Gamma_{n-1} ^*} \Gamma_n \\
0            & D_{\Gamma_2}           & \cdots & -\Gamma_2^* D_{\Gamma_3 ^*} \cdots D_{\Gamma_{n-1} ^*} \Gamma_n \\
\vdots       & \vdots                 & \ddots & \vdots\\
0 & 0 & \cdots & D_{\Gamma_n} \\
\end{array}
\right]
\]

\noindent and
\[
D_{T^*}^2 = D_{\Gamma_1 ^*}  \cdots D_{\Gamma_n ^*}  D_{\Gamma_n ^*}  \cdots  D_{\Gamma_1 ^*}. 
\]

\end{thm}

\pf The argument is by induction. The length $n = 2$ case was shown above. Now suppose the claim holds for length $n-1$. For a row contraction $T$
of length $n$, put

\[
T = [S, T_n]
\]

\noi where by inductive hypothesis,

\[
S = [\Gamma_1, D_{\Gamma_1 ^*} \Gamma_2, D_{\Gamma_1 ^*} D_{\Gamma_2 ^*} \Gamma_3, \cdots, D_{\Gamma_1 ^*} \cdots \Gamma_{n-1}].
\]

\noi According to lemma, there exists a contraction $\Lambda$ such that

\[
T_n = D_{S^*} \Lambda.
\]

\noi But 

\[
D_{S^*} = D_{\Gamma_1 ^*}  \cdots D_{\Gamma_n ^*} V
\]

\noi for some partial isometry $V$. Choosing $\Gamma_n = V \Lambda$ shows $T$ is of the desired form. Applying the defect operator identity 
proves the remaining proves the remaining claims. 
$\Box$

The combinatorial content of the theorem can be depicted pictorially. Figure 2 below shows the parametrization of length 3 row contractions. 
The downward arrows indicate input ports and the upward arrows output ports. For example, for a matrix $(T_{ij})$, each path from input $3$ to output $1$ 
gives rise to a term in the expression of the $T_{13}$ entry (in this particular case, $T_{13}$ is the third entry of a row contraction $T$).\\

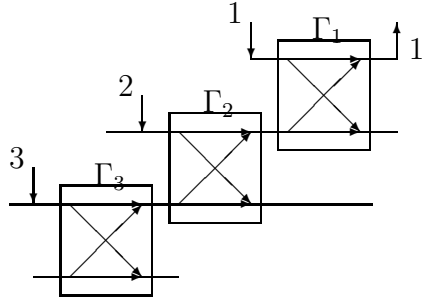
\begin{figure}[h]
\setlength{\unitlength}{3000sp}%
\begingroup\makeatletter\ifx\SetFigFont\undefined%
\gdef\SetFigFont#1#2#3#4#5{%
  \reset@font\fontsize{#1}{#2pt}%
  \fontfamily{#3}\fontseries{#4}\fontshape{#5}%
  \selectfont}%
\fi\endgroup%
\begin{picture}(6174,2949)(289,-2323)

%upward output arrow
{ \put(5701,-361){\vector( 0,1){300}}
}%

{ \put(5801,-361){$1$}
}%

%lower line connecting Box 2 and Box 3
{ \put(3301,-961){\line( 1, 0){1200}}
}%

%vector on the above line
{ \put(3901,-961){\vector( 1, 0){600}}
}%

{ \put(5000,-190){$\Gamma_1$}
}%

{ \put(4500,-61){\vector(0,-1){300}}
}%

{ \put(4300,-61){$1$}
}%

%Box 3
{ \put(4726,-1111){\framebox(750,900){}}
}%
{ \put(4501,-961){\line( 1, 0){1200}}
}%
{ \put(4501,-361){\line( 1, 0){1200}}
}%
{ \put(4801,-361){\vector( 1, 0){600}}
}%
{ \put(4801,-961){\vector( 1, 0){600}}
}%
{ \put(4801,-361){\vector( 1,-1){600}}
}%
{ \put(4801,-961){\vector( 1, 1){600}}
}%

{ \put(3600,-661){\vector(0,-1){300}}
}%
{ \put(3400,-661){$2$}
}%
{ \put(4100,-800){$\Gamma_2$}
}%

%Box5
{ \put(3826,-1711){\framebox(750,900){}}
}%
{ \put(3901,-961){\vector( 1,-1){600}}
}%
{ \put(3901,-1561){\vector( 1, 0){600}}
}%
{ \put(3901,-1561){\vector( 1, 1){600}}
}%

%lower line connecting Box 4 and Box 5
{ \put(2501,-1561){\line( 1, 0){3000}}
}%

{ \put(3200,-1400){$\Gamma_3$}
}%
{ \put(2700,-1261){\vector(0,-1){300}}
}%
{ \put(2500,-1261){$3$}
}%

%Box6
{ \put(2926,-2311){\framebox(750,900){}}
}%
{ \put(3001,-1561){\vector( 1, 0){600}}
}%
{ \put(3001,-2161){\vector( 1, 0){600}}
}%
{ \put(3001,-2161){\vector( 1, 1){600}}
}%
{ \put(3001,-1561){\vector( 1,-1){600}}
}%
{ \put(2701,-2161){\line( 1, 0){1200}}
}%

\end{picture}

\caption{Structure of row contractions of length 3}
\end{figure}

Figures 3 and 4 describe the parametrizations of the defect operators (more precisely that of the natural square roots, or Cholesky factors).
In the special case that $T T^* = I_{\cH}$, i.e. $T$ is a surjective partial isometry, the contraction $\Gamma_n$ is in fact a partial isometry.\\

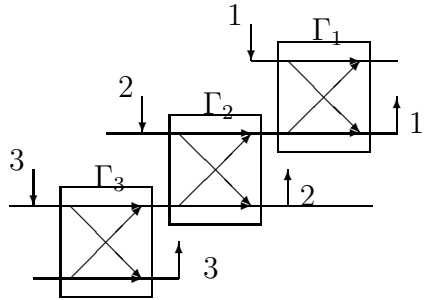
\begin{figure}[h]
\setlength{\unitlength}{3000sp}%
\begingroup\makeatletter\ifx\SetFigFont\undefined%
\gdef\SetFigFont#1#2#3#4#5{%
  \reset@font\fontsize{#1}{#2pt}%
  \fontfamily{#3}\fontseries{#4}\fontshape{#5}%
  \selectfont}%
\fi\endgroup%
\begin{picture}(6174,2949)(289,-2323)

%lower line connecting Box 2 and Box 3
{ \put(3301,-961){\line( 1, 0){1200}}
}%

%vector on the above line
{ \put(3901,-961){\vector( 1, 0){600}}
}%

{ \put(5000,-190){$\Gamma_1$}
}%

%first input downward arrow
{ \put(4300,-61){$1$}
}%
{ \put(4500,-61){\vector(0,-1){300}}
}%

%Box 3
{ \put(4726,-1111){\framebox(750,900){}}
}%
{ \put(4501,-961){\line( 1, 0){1200}}
}%
{ \put(4501,-361){\line( 1, 0){1200}}
}%
{ \put(4801,-361){\vector( 1, 0){600}}
}%
{ \put(4801,-961){\vector( 1, 0){600}}
}%
{ \put(4801,-361){\vector( 1,-1){600}}
}%
{ \put(4801,-961){\vector( 1, 1){600}}
}%

%first output upward arrow 
{ \put(5701,-961){\vector( 0,1){300}}
}%
{ \put(5801,-961){$1$}
}%

%second input downward arrow
{ \put(3600,-661){\vector(0,-1){300}}
}%
{ \put(3400,-661){$2$}
}%

{ \put(4100,-800){$\Gamma_2$}
}%

%Box5
{ \put(3826,-1711){\framebox(750,900){}}
}%
{ \put(3901,-961){\vector( 1,-1){600}}
}%
{ \put(3901,-1561){\vector( 1, 0){600}}
}%
{ \put(3901,-1561){\vector( 1, 1){600}}
}%

%second output upward arrow 
{ \put(4801,-1561){\vector( 0,1){300}}
}%
{ \put(4901,-1561){$2$}
}%

%lower line connecting Box 4 and Box 5
{ \put(2501,-1561){\line( 1, 0){3000}}
}%

%Third input arrow
{ \put(2700,-1261){\vector(0,-1){300}}
}%
{ \put(2500,-1261){$3$}
}%

{ \put(3200,-1400){$\Gamma_3$}
}%

%Box6
{ \put(2926,-2311){\framebox(750,900){}}
}%
{ \put(3001,-1561){\vector( 1, 0){600}}
}%
{ \put(3001,-2161){\vector( 1, 0){600}}
}%
{ \put(3001,-2161){\vector( 1, 1){600}}
}%
{ \put(3001,-1561){\vector( 1,-1){600}}
}%
{ \put(2701,-2161){\line( 1, 0){1200}}
}%

%Third output arrow
{ \put(3901,-2161){\vector( 0,1){300}}
}%
{ \put(4101,-2161){$3$}
}%

\end{picture}

\caption{Structure of $D_T$ where $T$ is a row contraction of length 3}
\end{figure}

\begin{figure}[h]
\setlength{\unitlength}{3000sp}%
\begingroup\makeatletter\ifx\SetFigFont\undefined%
\gdef\SetFigFont#1#2#3#4#5{%
  \reset@font\fontsize{#1}{#2pt}%
  \fontfamily{#3}\fontseries{#4}\fontshape{#5}%
  \selectfont}%
\fi\endgroup%
\begin{picture}(6174,2949)(289,-2323)

%upward output arrow
{ \put(5701,-361){\vector( 0,1){300}}
}%

{ \put(5801,-361){$1$}
}%

%lower line connecting Box 2 and Box 3
{ \put(3301,-961){\line( 1, 0){1200}}
}%

%vector on the above line
{ \put(3901,-961){\vector( 1, 0){600}}
}%

{ \put(5000,-190){$\Gamma_1$}
}%

%Box 3
{ \put(4726,-1111){\framebox(750,900){}}
}%
{ \put(4501,-961){\line( 1, 0){1200}}
}%
{ \put(4501,-361){\line( 1, 0){1200}}
}%
{ \put(4801,-361){\vector( 1, 0){600}}
}%
{ \put(4801,-961){\vector( 1, 0){600}}
}%
{ \put(4801,-361){\vector( 1,-1){600}}
}%
{ \put(4801,-961){\vector( 1, 1){600}}
}%

{ \put(4100,-800){$\Gamma_2$}
}%

%Box5
{ \put(3826,-1711){\framebox(750,900){}}
}%
{ \put(3901,-961){\vector( 1,-1){600}}
}%
{ \put(3901,-1561){\vector( 1, 0){600}}
}%
{ \put(3901,-1561){\vector( 1, 1){600}}
}%

%lower line connecting Box 4 and Box 5
{ \put(2501,-1561){\line( 1, 0){3000}}
}%

{ \put(3200,-1400){$\Gamma_3$}
}%

%Box6
{ \put(2926,-2311){\framebox(750,900){}}
}%
{ \put(3001,-1561){\vector( 1, 0){600}}
}%
{ \put(3001,-2161){\vector( 1, 0){600}}
}%
{ \put(3001,-2161){\vector( 1, 1){600}}
}%
{ \put(3001,-1561){\vector( 1,-1){600}}
}%
{ \put(2701,-2161){\line( 1, 0){1200}}
}%

%downward input arrow
{ \put(2701,-1861){\vector(0,-1){300}}
}%

{ \put(2501,-1861){$1$}
}%

\end{picture}

\caption{Structure of $D_T^*$ where $T$ is a row contractions of length 3}
\end{figure}
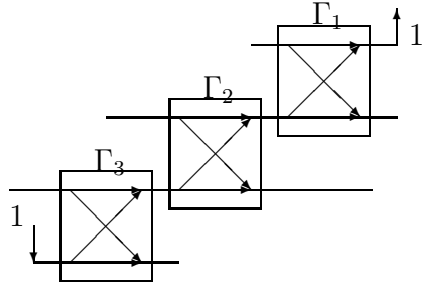

The structure of column contractions can be exhibited in similar fashion. For completeness, we state the corresponding result
for column contractions.

\begin{thm} An operator

\[
T = 
\left[
\begin{array}{c}
T_1 \\ \vdots \\T_n
\end{array} \right]
: \cH \rightarrow \oplus _1 ^n \cH_i
\]

\noi is a contraction if and only if

\[
T_1 = \Gamma_1 \quad \mbox{and} \quad T_k = \Gamma_k D_{\Gamma_{k-1}} \cdots D_{\Gamma_1},
\]

\noi where $\Gamma_i$'s are contractions. Moreoever, the defect operators of $T$ and $T^*$ take the form

\[
D_T ^2 = D_{\Gamma_1}  \cdots D_{\Gamma_n}  D_{\Gamma_n}  \cdots  D_{\Gamma_1}, 
\]

\noi and

\[
D_{T^*}^2
=
\left[
\begin{array}{cccc}
D_{\Gamma_1 ^*}                                                     & 0               & \cdots & 0 \\
- \Gamma_2 \Gamma_1 ^*                                             & D_{\Gamma_2 ^*}  & \cdots &  0         \\
\vdots                                                             & \vdots           & \ddots & \vdots\\
-\Gamma_n D_{\Gamma_{n-1} } \cdots D_{\Gamma_2 } \Gamma_1^*    & -\Gamma_n D_{\Gamma_{n-1} } \cdots D_{\Gamma_{2} } \Gamma_2^* & \cdots & D_{\Gamma_n^*} \\
\end{array}
\right]
\left[
\begin{array}{cccc}
D_{\Gamma_1 ^*} & - \Gamma_1 \Gamma_2^* & \cdots & -\Gamma_1 D_{\Gamma_2} \cdots D_{\Gamma_{n-1}} \Gamma_n ^* \\
0            & D_{\Gamma_2 ^*}          & \cdots & -\Gamma_2 D_{\Gamma_3 } \cdots D_{\Gamma_{n-1} } \Gamma_n ^* \\
\vdots       & \vdots                   & \ddots & \vdots\\
0 & 0 & \cdots & D_{\Gamma_n ^*} \\
\end{array}
\right]
\]

\end{thm}

One can be easily convinced that there are lattice diagrams corresponding to the above description. If a column contraction $T$ is such that
$T^* T = I_{\cH}$ (i.e. $T$ is an isometry from $\cH$ to $\oplus_1 ^n \cH_i$), $\Gamma_n$ is a partial isometry.

\section{Matrix contractions}

Let $T: \cH_1 \oplus \cH_2 \rightarrow \cK_1 \oplus \cK_2$,
 
\[
T = 
\left[
\begin{array}{cc}
A & B \\
C & D \\
\end{array}
\right],
\] 

\noi be a contraction. Then $[A \; B] = P_{\cK_1} T$, where $P_{\cK_1}$ denotes projection onto $\cK_1$, is necessarily a row contraction, therefore
of the form $[\Gamma_1 \; D_{\Gamma_1 ^*} \Gamma_2]$ where $\Gamma_i$ are contractions. Similarly, we have

\[
\left[
\begin{array}{c} A \\ C \end{array} 
\right]
=  
\left[
\begin{array}{c} \Gamma_1 \\ \Gamma_3 D_{\Gamma_1} \end{array}
\right]. 
\]

\noi So we have

\[
T = 
\left[
\begin{array}{cc}
\Gamma_1 & D_{\Gamma_1 ^*} \Gamma_2\\
\Gamma_3 D_{\Gamma_1}  & D \\
\end{array}
\right].
\]

\noi We will show that the entry $D$, unspecified so far, can also be parametrized by contractions. To this end, view $T$ as a row contraction
$T = [S_1 \; S_2]$ with 

\[
S_1 = 
\left[
\begin{array}{c} \Gamma_1 \\ \Gamma_3 D_{\Gamma_1} \end{array}
\right]
\quad \mbox{and} \quad
S_2 = 
\left[
\begin{array}{c} D_{\Gamma_1^*} \Gamma_2 \\ D \end{array}
\right].
\]

\noi Therefore $S_2 = D_{S_1^*} \Lambda$ for some (column) contraction $\Lambda$. $S_1$ is a column contraction and direct calculation gives

\[
D_{S_1}^2 = 
\left[
\begin{array}{cc}
D_{\Gamma_1 ^*} & 0\\
- \Gamma_3 \Gamma_1^*  & D_{\Gamma_3} \\
\end{array}
\right]
\left[
\begin{array}{cc}
D_{\Gamma_1 ^*} & - \Gamma_1 \Gamma_3^*\\
0  & D_{\Gamma_3}\\
\end{array}
\right].
\]

\noi Thus

\[
S_2 = 
\left[
\begin{array}{cc}
D_{\Gamma_1 ^*} & 0\\
- \Gamma_3 \Gamma_1^*  & D_{\Gamma_3} \\
\end{array}
\right]
\left[
\begin{array}{c} \Lambda_1 \\ \Lambda_2 D_{\Lambda_1} \end{array}
\right].
\]

\noi Comparing entries and invoking the uniqueness condition shows that $\Lambda_1 = \Gamma_2$. Rename $\Lambda_2$ as $\Gamma_4$ and we have

\[
D = - \Gamma_3 \Gamma_1^* \Gamma_2 + D_{\Gamma_3} \Gamma_4 D_{\Gamma_2}.
\]

The above can be summarized by \cite{Gheon}:

\begin{thm}Every $2 \times 2$ contraction $T: \cH_1 \oplus \cH_2 \rightarrow \cK_1 \oplus \cK_2$ is of the form 

\[
T = 
\left[
\begin{array}{cc}
\Gamma_1 & D_{\Gamma_1 ^*} \Gamma_2\\
\Gamma_3 D_{\Gamma_1}  & - \Gamma_3 \Gamma_1^* \Gamma_2 + D_{\Gamma_3} \Gamma_4 D_{\Gamma_2} \\
\end{array}
\right]
\]

\noi where $\Gamma_i$ are contractions.
\end{thm}

The general structure of a $n \times m$ matrix contraction can be obtained in a similar way.

\begin{thm} Let

\[
T = 
\left[
\begin{array}{ccc}
T_{11} & \cdots & T_{1m} \\
\vdots & \ddots & \vdots\\
T_{n1} & \cdots & T_{nm} 
\end{array}
\right]
:
\oplus_1 ^m \cH_i \rightarrow \oplus_1 ^n \cK_i
\]

\noi be a contraction. Then the column contraction 

\[
C_1
=
\left[ \begin{array}{c}
T_{11}\\
\vdots\\
T_{n1} 
\end{array} \right]
\quad
\mbox{is of the form}
\quad
\left[ \begin{array}{c}
\Gamma_1\\
\Gamma_2 D_{\Gamma_1}\\
\vdots\\
\Gamma_n D_{\Gamma_{n-1}} \cdots D_{\Gamma_1} 
\end{array} \right]
\quad
\mbox{where $\Gamma_i$ are contractions}.
\]

\noi For $1 < k \leq m $, the column 

\[
\left[ \begin{array}{c}
T_{1k}\\
\vdots\\
T_{nk} 
\end{array} \right]
\quad
\mbox{is obtained inductively by}
\quad
\left[ \begin{array}{c}
T_{1k}\\
\vdots\\
T_{nk} 
\end{array} \right] 
=
D_{C_1 ^*} D_{C_2 ^*} \cdots D_{C_{k-1}^*} C_k 
\]

\noi where $C_i$ is the column contraction parametrized by $\Gamma_{n(i - 1) + 1}$...$\Gamma_{ni}$ with $C_1$ being as specified above.
\end{thm}

\noi The proof is immediate and omitted. The combinatorial structure of matrix contractions can also be nicely described via lattice paths. 
For example, the lattice diagram for the $2 \times 2$ case is figure 5.

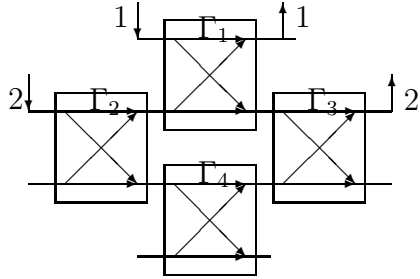
\begin{figure}[h]
\setlength{\unitlength}{3000sp}%
\begingroup\makeatletter\ifx\SetFigFont\undefined%

\gdef\SetFigFont#1#2#3#4#5{%
  \reset@font\fontsize{#1}{#2pt}%
  \fontfamily{#3}\fontseries{#4}\fontshape{#5}%
  \selectfont}%
\fi\endgroup%

\begin{picture}(6174,2949)(289,-2323)

%input arrow 1
{ \put(2701,-61){\vector(0, -1){300}}
}%
{ \put(2501,-261){$1$}
}%

%upper line to the left of Box 2
{ \put(2701,-361){\line( 1, 0){400}}
}%

%lower line connecting Box 1 and Box 2
{ \put(2101,-961){\line( 1, 0){1200}}
}%

% the vector on the above line
{ \put(2101,-961){\vector( 1, 0){600}}
}%

%Box2
{ \put(3201,-351){$\Gamma_1$}
}%
{ \put(2926,-1111){\framebox(750,900){}}
}%
{ \put(3001,-361){\vector( 1, 0){600}}
}%
{ \put(3001,-961){\vector( 1, 0){600}}
}%
{ \put(3001,-961){\vector( 1, 1){600}}
}%
{ \put(3001,-361){\vector( 1,-1){600}}
}%

%upper line connecting Box 2 and Box 3
{ \put(3601,-361){\line( 1, 0){400}}
}%

%output arrow 1
{ \put(3901,-361){\vector(0, 1){300}}
}%
{ \put(4001,-261){$1$}
}%

%lower line connecting Box 2 and Box 3
{ \put(3301,-961){\line( 1, 0){1200}}
}%

%vector on the above line
{ \put(3901,-961){\vector( 1, 0){600}}
}%

%upper line to the left of Box 4
{ \put(1801,-961){\line( 1, 0){3000}}
}%

%input arrow 2
{ \put(1801,-661){\vector(0, -1){300}}
}%
{ \put(1631,-941){$2$}
}%

%Box4
{ \put(2301,-951){$\Gamma_2$}
}%
{ \put(2026,-1711){\framebox(750,900){}}
}%
{ \put(2101,-1561){\vector( 1, 0){600}}
}%
{ \put(2101,-1561){\vector( 1, 1){600}}
}%
{ \put(2101,-961){\vector( 1,-1){600}}
}%

%lower line connecting Box 4 and Box 5
{ \put(1801,-1561){\line( 1, 0){3000}}
}%

%Box5
{ \put(4101,-951){$\Gamma_3$}
}%
{ \put(3826,-1711){\framebox(750,900){}}
}%
{ \put(3901,-961){\vector( 1,-1){600}}
}%
{ \put(3901,-1561){\vector( 1, 0){600}}
}%
{ \put(3901,-1561){\vector( 1, 1){600}}
}%

%output arrow 2
{ \put(4801,-961){\vector( 0, 1){300}}
}%
{ \put(4901,-951){$2$}
}%

%Box6
{ \put(3201,-1551){$\Gamma_4$}
}%
{ \put(2926,-2311){\framebox(750,900){}}
}%
{ \put(3001,-1561){\vector( 1, 0){600}}
}%
{ \put(3001,-2161){\vector( 1, 0){600}}
}%
{ \put(3001,-2161){\vector( 1, 1){600}}
}%
{ \put(3001,-1561){\vector( 1,-1){600}}
}%

%lower line
{ \put(2701,-2161){\line( 1, 0){1100}}
}%

\end{picture}

\caption{A $2 \times 2$ matrix contraction}
\end{figure}

The defect operators of matrix contractions can also be calculated. Due to the "two-layer" nature of its parametrization, the explicit formulae may
seem complicated. It is helpful to first look at the lattice diagrams. From inspecting the above figure, we anticipate that the Cholesky factor 
of $D_T$ and $D_{T^*}$ to have the corresponding pictures given by figures 6 and 7 respectively.

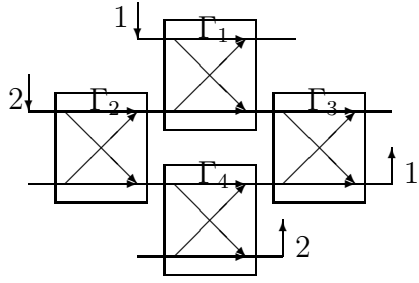
\begin{figure}[h]
\setlength{\unitlength}{3000sp}%
\begingroup\makeatletter\ifx\SetFigFont\undefined%

\gdef\SetFigFont#1#2#3#4#5{%
  \reset@font\fontsize{#1}{#2pt}%
  \fontfamily{#3}\fontseries{#4}\fontshape{#5}%
  \selectfont}%
\fi\endgroup%

\begin{picture}(6174,2949)(289,-2323)

%input arrow 1
{ \put(2701,-61){\vector(0, -1){300}}
}%
{ \put(2501,-261){$1$}
}%

%upper line to the left of Box 2
{ \put(2701,-361){\line( 1, 0){400}}
}%

%lower line connecting Box 1 and Box 2
{ \put(2101,-961){\line( 1, 0){1200}}
}%

% the vector on the above line
{ \put(2101,-961){\vector( 1, 0){600}}
}%

%Box2
{ \put(3201,-351){$\Gamma_1$}
}%
{ \put(2926,-1111){\framebox(750,900){}}
}%
{ \put(3001,-361){\vector( 1, 0){600}}
}%
{ \put(3001,-961){\vector( 1, 0){600}}
}%
{ \put(3001,-961){\vector( 1, 1){600}}
}%
{ \put(3001,-361){\vector( 1,-1){600}}
}%

%upper line connecting Box 2 and Box 3
{ \put(3601,-361){\line( 1, 0){400}}
}%

%lower line connecting Box 2 and Box 3
{ \put(3301,-961){\line( 1, 0){1200}}
}%

%vector on the above line
{ \put(3901,-961){\vector( 1, 0){600}}
}%

%upper line to the left of Box 4
{ \put(1801,-961){\line( 1, 0){3000}}
}%

%input arrow 2
{ \put(1801,-661){\vector(0, -1){300}}
}%
{ \put(1631,-941){$2$}
}%

%Box4
{ \put(2301,-951){$\Gamma_2$}
}%
{ \put(2026,-1711){\framebox(750,900){}}
}%
{ \put(2101,-1561){\vector( 1, 0){600}}
}%
{ \put(2101,-1561){\vector( 1, 1){600}}
}%
{ \put(2101,-961){\vector( 1,-1){600}}
}%

%lower line connecting Box 4 and Box 5
{ \put(1801,-1561){\line( 1, 0){3000}}
}%

%Box5
{ \put(4101,-951){$\Gamma_3$}
}%
{ \put(3826,-1711){\framebox(750,900){}}
}%
{ \put(3901,-961){\vector( 1,-1){600}}
}%
{ \put(3901,-1561){\vector( 1, 0){600}}
}%
{ \put(3901,-1561){\vector( 1, 1){600}}
}%

%output arrow 1
{ \put(4801,-1561){\vector( 0, 1){300}}
}%
{ \put(4901,-1551){$1$}
}%

%Box6
{ \put(3201,-1551){$\Gamma_4$}
}%
{ \put(2926,-2311){\framebox(750,900){}}
}%
{ \put(3001,-1561){\vector( 1, 0){600}}
}%
{ \put(3001,-2161){\vector( 1, 0){600}}
}%
{ \put(3001,-2161){\vector( 1, 1){600}}
}%
{ \put(3001,-1561){\vector( 1,-1){600}}
}%

%lower line
{ \put(2701,-2161){\line( 1, 0){1200}}
}%

%output arrow 2
{ \put(3901,-2161){\vector(0, 1){300}}
}%
{ \put(4001,-2161){$2$}
}%

\end{picture}

\caption{The natural Cholesky factor of $D_T$ where $T$ is a $2 \times 2$ matrix contraction}
\end{figure}

\begin{figure}[h]
\setlength{\unitlength}{3000sp}%
\begingroup\makeatletter\ifx\SetFigFont\undefined%

\gdef\SetFigFont#1#2#3#4#5{%
  \reset@font\fontsize{#1}{#2pt}%
  \fontfamily{#3}\fontseries{#4}\fontshape{#5}%
  \selectfont}%
\fi\endgroup%

\begin{picture}(6174,2949)(289,-2323)

%upper line to the left of Box 2
{ \put(2701,-361){\line( 1, 0){400}}
}%

%lower line connecting Box 1 and Box 2
{ \put(2101,-961){\line( 1, 0){1200}}
}%

% the vector on the above line
{ \put(2101,-961){\vector( 1, 0){600}}
}%

%Box2
{ \put(3201,-351){$\Gamma_1$}
}%
{ \put(2926,-1111){\framebox(750,900){}}
}%
{ \put(3001,-361){\vector( 1, 0){600}}
}%
{ \put(3001,-961){\vector( 1, 0){600}}
}%
{ \put(3001,-961){\vector( 1, 1){600}}
}%
{ \put(3001,-361){\vector( 1,-1){600}}
}%

%upper line connecting Box 2 and Box 3
{ \put(3601,-361){\line( 1, 0){400}}
}%

%output arrow 1
{ \put(3901,-361){\vector(0, 1){300}}
}%
{ \put(4001,-261){$1$}
}%

%lower line connecting Box 2 and Box 3
{ \put(3301,-961){\line( 1, 0){1200}}
}%

%vector on the above line
{ \put(3901,-961){\vector( 1, 0){600}}
}%

%upper line to the left of Box 4
{ \put(1801,-961){\line( 1, 0){3000}}
}%

%input arrow 1
{ \put(1801,-1261){\vector(0, -1){300}}
}%
{ \put(1631,-1441){$1$}
}%

%Box4
{ \put(2301,-951){$\Gamma_2$}
}%
{ \put(2026,-1711){\framebox(750,900){}}
}%
{ \put(2101,-1561){\vector( 1, 0){600}}
}%
{ \put(2101,-1561){\vector( 1, 1){600}}
}%
{ \put(2101,-961){\vector( 1,-1){600}}
}%

%lower line connecting Box 4 and Box 5
{ \put(1801,-1561){\line( 1, 0){3000}}
}%

%Box5
{ \put(4101,-951){$\Gamma_3$}
}%
{ \put(3826,-1711){\framebox(750,900){}}
}%
{ \put(3901,-961){\vector( 1,-1){600}}
}%
{ \put(3901,-1561){\vector( 1, 0){600}}
}%
{ \put(3901,-1561){\vector( 1, 1){600}}
}%

%output arrow 2
{ \put(4801,-961){\vector( 0, 1){300}}
}%
{ \put(4901,-951){$2$}
}%

%input arrow 2
{ \put(2701,-1861){\vector( 0, -1){300}}
}%
{ \put(2551,-2051){$2$}
}%

%Box6
{ \put(3201,-1551){$\Gamma_4$}
}%
{ \put(2926,-2311){\framebox(750,900){}}
}%
{ \put(3001,-1561){\vector( 1, 0){600}}
}%
{ \put(3001,-2161){\vector( 1, 0){600}}
}%
{ \put(3001,-2161){\vector( 1, 1){600}}
}%
{ \put(3001,-1561){\vector( 1,-1){600}}
}%

%lower line
{ \put(2701,-2161){\line( 1, 0){1100}}
}%

\end{picture}

\caption{The Cholesky factor for $D_{T^*}$}
\end{figure}
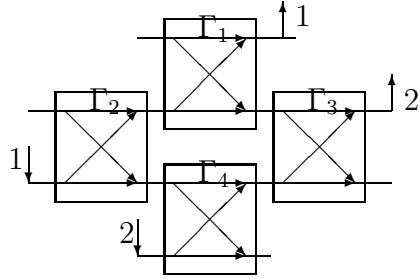

In other words, one should have

\[
D_T ^2 =
\left[
\begin{array}{cc}
D_{\Gamma_3} D_{\Gamma_1} & 0\\
- \Gamma_2^* \Gamma_1 D_{\Gamma_3} - D_{\Gamma_2} \Gamma_4 ^* \Gamma_3 & D_{\Gamma_2} D_{\Gamma_4}
\end{array} \right]
 \left[
\begin{array}{cc}
 D_{\Gamma_3} D_{\Gamma_1} & -D_{\Gamma_3} \Gamma_1 ^* \Gamma_2 - \Gamma_3 ^* \Gamma_4 D_{\Gamma_2} \\
0 & D_{\Gamma_4} D_{\Gamma_2}
\end{array} \right],
\]

\noi and

\[
D_{T^*}^2 = 
\left[
\begin{array}{cc}
D_{\Gamma_1^*} D_{\Gamma_2 ^*} & 0\\
- \Gamma_2 \Gamma_4 ^* D_{\Gamma_3^*} - D_{\Gamma_2 ^*} \Gamma_1 ^* \Gamma_3^* & D_{\Gamma_4^*} D_{\Gamma_2^*}
\end{array} \right]
\left[
\begin{array}{cc}
 D_{\Gamma_2^*} D_{\Gamma_1^*} & -D_{\Gamma_3^*} \Gamma_4 \Gamma_2 ^* - \Gamma_3 \Gamma_1 D_{\Gamma_2^*} \\
0 & D_{\Gamma_2 ^*} D_{\Gamma_4 ^*}
\end{array} \right].
\]

\noi This can be confirmed by a straightforward but perhaps tedious calculation, which we shall not bore the reader with. The defect operators for matrix
contraction of any finite size can obtained in similar fashion.\\

\noi
{\em Unitary Matrices} The unitary operators are the extreme points of contractions, thus a special case. If

\[
T = 
\left[
\begin{array}{cc}
\Gamma_1 & D_{\Gamma_1 ^*} \Gamma_2\\
\Gamma_3 D_{\Gamma_1}  & - \Gamma_3 \Gamma_1^* \Gamma_2 + D_{\Gamma_3} \Gamma_4 D_{\Gamma_2} \\
\end{array}
\right]
: \cH_1 \oplus \cH_2 \rightarrow \cK_1 \oplus \cK_2
\] 

\noi
is unitary, then $[\Gamma_1 \; D_{\Gamma_1 ^*} \Gamma_2]$ is a partial isometry therefore so is $\Gamma_2$; same goes for $\Gamma_3$. When all
spaces are finite dimensional and, in the expression

\[
T = 
\left[
\begin{array}{cc}
A & B\\
C  & D\\
\end{array}
\right],
\] 

\noi we have $B$ and $C$ being square matrices, $\Gamma_2$ and $\Gamma_3$ are unitary and the description of T becomes very simple:

\[
T = 
\left[
\begin{array}{cc}
\Gamma_1 & D_{\Gamma_1 ^*} \Gamma_2\\
\Gamma_3 D_{\Gamma_1}  & - \Gamma_3 \Gamma_1^* \Gamma_2\\
\end{array}
\right].
\] 

\noi In other words, all unitary matrice are related to the Julia operator via

\[
T = 
\left[
\begin{array}{cc}
1 & 0\\
0 & \Gamma_3\\
\end{array}
\right]
\left[
\begin{array}{cc}
\Gamma_1 & D_{\Gamma_1 ^*}\\
D_{\Gamma_1}  & - \Gamma_1^*\\
\end{array}
\right]
\left[
\begin{array}{cc}
1 & 0\\
0  & \Gamma_2\\
\end{array}
\right].
\] 

\noi This will be applied in the sequel in calculating dilations of completely positive maps/quantum operations.

\section{Positive Matrices}

Similar dilation-theoretic techniques can be applied to positive matrices. As stated in the introduction, we outline basic results for completeness. 
See \cite{Co} for a comprehensive discussion. As for contractions, one can start by examining the $2 \times 2$ case then apply induction. Let

\[
A = 
\left[ 
\begin{array}{cc}
L_1 ^* L_1 & A_{12} \\
A_{12}^* & L_2 ^* L_2
\end{array}
\right]
\in {\mathcal L}( {\mathcal H}_1 \oplus {\mathcal H} _2 )
\] 

\noindent
be a positive semidefinite operator matrix whose entries are bounded operators, that is 

\[
\langle \left[ \begin{array}{c} h_1 \\ h_2\end{array} \right], A  \left[\begin{array}{c} h_1 \\ h_2\end{array}\right] \rangle 
_{{\mathcal H}_1 \oplus {\mathcal H}_2} 
\geq 0
\] 

\noindent
for all $h_1 \in {\mathcal H} _1$ and $h_2 \in {\mathcal H} _2$.

\begin{thm}
There exists an unique contraction $\Gamma \in {\mathcal L}({\mathcal H}_2 , {\mathcal H}_1)$ such that 
$A_{12} = L_1^* \Gamma L_2$.
\end{thm}

\pf: Assume for the moment that both $A_{11}$ and $A_{22}$ are invertible. 
Then a {\sl Frobenius-Schur identity} holds:

\[
A = 
\left[ 
\begin{array}{cc}
L_1 ^* L_1 & A_{12} \\
A_{12}^* & L_2 ^* L_2
\end{array}
\right]
=
\left[ 
\begin{array}{cc}
I & 0 \\
A_{12}^* (L_1 ^* L_1)^{-1} & I
\end{array}
\right]
\cdot
\left[ 
\begin{array}{cc}
L_1 ^* L_1 & 0\\
0 & L_2 ^* L_2 - A_{12} ^* (L_1 ^* L_1)^{-1} A_{12}
\end{array}
\right]
\cdot
\left[ 
\begin{array}{cc}
I & (L_1 ^* L_1) ^{-1} A_{12} \\
0 & I
\end{array}
\right].
\]

\noindent It follows that $A$ is positive if and only if its {\sl Schur complement} 

\[
L_2 ^* L_2 - A_{12} ^* (L_1 ^* L_1)^{-1} A_{12} \geq 0 
\]

\noindent i.e.

\[
L_2 ^* L_2  \geq A_{12} ^* (L_1)^{-1} (L_1 ^*)^{-1} A_{12}.
\]

\noindent By lemma 1, there exist a contraction $\Gamma$ such that $\Gamma L_2  = (L_1 ^{-1})^* A_{12}$, where we impose the condition required for 
uniqueness. Thus $A_{12} = L_1 ^* \Gamma L_2$.

For the general case where $A_{11} = L_1 ^* L_1$ and $A_{22} = L_2 ^* L_2$ need not be invertible, consider the sequences 
$\{ \alpha_n ^* \alpha_n = A_{11} + \frac{1}{n} \}$ and $\{\beta _n ^* \beta_n = A_{22} + \frac{1}{n} \}$. By the spectral mapping theorem for 
self adjoint operators, $\alpha_n ^* \alpha_n$ and $\beta _n ^* \beta_n$ are invertible for all $n$. Therefore there exist 
contractions $\{ \Gamma _n\}$ with $A_{12} = \alpha _n ^* \Gamma _n \beta _n $. Since the unit ball in 
${\mathcal L}({\mathcal H}_2, {\mathcal H}_1)$ is compact in the weak operator topology, $\Gamma _n$ converges to some contraction $\Gamma$ weakly.
We can compute directly, for all $h_1 \in {\mathcal H} _1$ and $h_2 \in {\mathcal H} _2$,

\[
\langle h_1, L_1^* \Gamma L_2 h_2 \rangle _{ {\mathcal H}_1 }
= \lim_n \langle h_1, L_1^* \Gamma_n L_2 h_2 \rangle _{ {\mathcal H}_1 }
= \lim_n \langle h_1, \alpha_n ^* \Gamma_n \beta _n h_2 \rangle _{ {\mathcal H}_1 }
= \langle h_1, A_{12} h_2 \rangle _{ {\mathcal H}_1 }.
\]

\noi This proves the claim. $\Box$\\

This can be generalized to positive operator matrices of arbitrary size in the obvious way. We present the finite case as an algorithm

\begin{algo} \cite{Co}
Let $A = [ A_{ij} ]_{ij} \in {\mathcal L}(\oplus _{i=1} ^n {\mathcal H}_i)$ 
be positive. The {\bf Schur-Constantinescu}, or {\bf SC}, parametrization of $A$ can be calculated recursively
as follows:\\

i) $\left[ \begin{array}{cc} A_{n-1, n-1} & A_{n-1, n}\\ A_{n, n-1} & A_{n, n} \end{array} \right]$ is positive and can be parametrized 
according to 

the $2 \times 2$ case.

ii) For $1 \leq k \leq n-2$, the SC parametrization of
$\left[ \begin{array}{cccc} A_{k, k} & A_{k, k+1} & \cdots & A_{k,n}\\ A_{k+1, k} & A_{k+1, k+1}  & \cdots & A_{k+1, n}\\ \vdots & \vdots        
& \ddots & \vdots\\ A_{n, k} & A_{n, k+1}  & \cdots & A_{n,n} \end{array} \right]$ 

is calculated by first considering 
$\left[ \begin{array}{ccc} A_{k+1, k+1}  & \cdots & A_{k+1, n}\\ \vdots & \ddots & \vdots \\ A_{n, k+1} & \cdots 
& A_{n,n} \end{array} \right] = L_{k+1, k+1}^* L_{k+1, k+1}$, 

where $L_{k+1}$ is the Cholesky factor calculated in the previous step then put 

\[
[ A_{k, k+1} \; \cdots \; A_{k,n} ] = A_{k, k} ^{\frac{1}{2}} R_k L_{k+1}
\]
 
with $R_k$ being the corresponding row contraction.
\end{algo}

The lattice diagram for a $4 \times 4$ positive mattrix is given below.\\

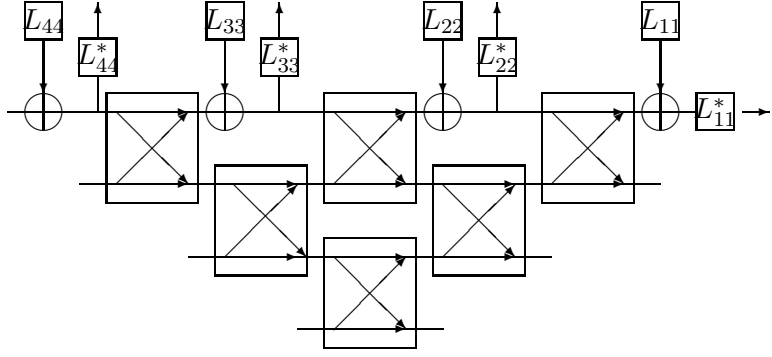
\begin{figure}[h]
\setlength{\unitlength}{3000sp}%
\begingroup\makeatletter\ifx\SetFigFont\undefined%
\gdef\SetFigFont#1#2#3#4#5{%
  \reset@font\fontsize{#1}{#2pt}%
  \fontfamily{#3}\fontseries{#4}\fontshape{#5}%
  \selectfont}%
\fi\endgroup%
\begin{picture}(6324,2874)(289,-2323)
{ \thinlines
\put(601,-361){\circle{300}}
}%
{ \put(2101,-361){\circle{300}}
}%
{ \put(3901,-361){\circle{300}}
}%
{ \put(5701,-361){\circle{300}}
}%
{ \put(301,-361){\line( 1, 0){900}}
}%
{ \put(601,-211){\line( 0,-1){300}}
}%
{ \put(1201,-361){\vector( 1,-1){600}}
}%
{ \put(1201,-961){\vector( 1, 1){600}}
}%
{ \put(451,239){\framebox(300,300){\small  $L_{44}$}}
}%
{ \put(601,239){\vector( 0,-1){450}}
}%
{ \put(1051,-361){\line( 0, 1){300}}
}%
{ \put(1051,239){\vector( 0, 1){300}}
}%
{ \put(901,-61){\framebox(300,300){\small $L^*_{44}$}}
}%
{ \put(1126,-1111){\framebox(750,900){}}
}%
{ \put(901,-961){\line( 1, 0){1200}}
}%
{ \put(1201,-361){\line( 1, 0){1200}}
}%
{ \put(2926,-1111){\framebox(750,900){}}
}%
{ \put(2026,-1711){\framebox(750,900){}}
}%
{ \put(3826,-1711){\framebox(750,900){}}
}%
{ \put(4726,-1111){\framebox(750,900){}}
}%
{ \put(2926,-2311){\framebox(750,900){}}
}%
{ \put(2101,-961){\line( 1, 0){1200}}
}%
{ \put(3301,-961){\line( 1, 0){1200}}
}%
{ \put(4501,-961){\line( 1, 0){1200}}
}%
{ \put(2401,-361){\line( 1, 0){1200}}
}%
{ \put(3601,-361){\line( 1, 0){1200}}
}%
{ \put(4801,-361){\line( 1, 0){1200}}
}%
{ \put(1201,-361){\vector( 1, 0){600}}
}%
{ \put(1201,-961){\vector( 1, 0){600}}
}%
{ \put(3001,-361){\vector( 1, 0){600}}
}%
{ \put(3001,-961){\vector( 1, 0){600}}
}%
{ \put(2101,-961){\vector( 1, 0){600}}
}%
{ \put(2101,-1561){\vector( 1, 0){600}}
}%
{ \put(3001,-1561){\vector( 1, 0){600}}
}%
{ \put(3001,-2161){\vector( 1, 0){600}}
}%
{ \put(3901,-961){\vector( 1, 0){600}}
}%
{ \put(3901,-1561){\vector( 1, 0){600}}
}%
{ \put(4801,-361){\vector( 1, 0){600}}
}%
{ \put(4801,-961){\vector( 1, 0){600}}
}%
{ \put(3001,-961){\vector( 1, 1){600}}
}%
{ \put(4801,-961){\vector( 1, 1){600}}
}%
{ \put(3001,-361){\vector( 1,-1){600}}
}%
{ \put(4801,-361){\vector( 1,-1){600}}
}%
{ \put(2101,-1561){\vector( 1, 1){600}}
}%
{ \put(3901,-1561){\vector( 1, 1){600}}
}%
{ \put(3001,-2161){\vector( 1, 1){600}}
}%
{ \put(2176,-961){\vector( 1,-1){600}}
}%
{ \put(3001,-1561){\vector( 1,-1){600}}
}%
{ \put(3901,-961){\vector( 1,-1){600}}
}%
{ \put(1801,-1561){\line( 1, 0){3000}}
}%
{ \put(2701,-2161){\line( 1, 0){1200}}
}%
{ \put(2101,-511){\line( 0, 1){300}}
}%
{ \put(3901,-511){\line( 0, 1){300}}
}%
{ \put(5701,-511){\line( 0, 1){300}}
}%
{ \put(1951,239){\framebox(300,300){\small $L_{33}$}}
}%
{ \put(3751,239){\framebox(300,300){\small $L_{22}$}}
}%
{ \put(5551,239){\framebox(300,300){\small $L_{11}$}}
}%
{ \put(2101,239){\vector( 0,-1){450}}
}%
{ \put(3901,239){\vector( 0,-1){450}}
}%
{ \put(5701,239){\vector( 0,-1){450}}
}%
{ \put(2401,-61){\framebox(300,300){\small $L^*_{33}$}}
}%
{ \put(4201,-61){\framebox(300,300){\small $L^*_{22}$}}
}%
{ \put(2551,-361){\line( 0, 1){300}}
}%
{ \put(4351,-361){\line( 0, 1){300}}
}%
{ \put(2551,239){\vector( 0, 1){300}}
}%
{ \put(4351,239){\vector( 0, 1){300}}
}%
{ \put(6001,-511){\framebox(300,300){\small $L^*_{11}$}}
}%
{ \put(6376,-361){\vector( 1, 0){225}}
}%
\end{picture}

\caption{ Lattice structure for $4\times 4$ positive matrices}
\end{figure}

\noi {\em Tensor product of positive matrices}
If $M = (m_{ij}) \in \bC^{n \times n}$ is a positive matrix with complex entries and $A = B^*B \in \cL(\cH)$ where $\cH$ is a Hilbert space.
Suppose $M$ is SC-parametrized by $\{ \gamma_i\} \subset \bC$, $i = 1, \cdots, \frac{1}{2}m(m-1)$ with $|\gamma_i| \leq 1$. The parametrization
of the positive matrix

\[
M \otimes A = (m_{ij} L^*L) \in \bC^{n \times n} \otimes \cL(\cH)
\]

\noi can be described easily \cite{tiviII}. Namely, let $\Gamma_i = \gamma_i I_{\cH}$ and $\L_i = \sqrt{m_{ii}} B$; it is clear that
they parametrize $M \otimes A$ in the sense of Schur-Constantinescu.\\

\noi {\em Matrices given by a strict inequality.} The natural square roots given by the SC-parametrization are 
upper(or lower)- triangular, i.e. they are Cholesky factors. Owing to this fact, if $A^* A \geq B^*B$, and the SC parameters of $A^*A$ are known, 
$B$ is readily described. Take for instance the $2 \times 2$ case. $B = \Gamma A$, where $\Gamma$ is a contraction. Let

\[
A =
\left[ 
\begin{array}{cc}
L_{11} & \Lambda L_{22} \\
0 & D_{\Lambda} L_{22}
\end{array} \right]
\quad \mbox{and} \quad
\Gamma = 
\left[
\begin{array}{cc}
\Gamma_1 & D_{\Gamma_1 ^*} \Gamma_2\\
\Gamma_3 D_{\Gamma_1}  & - \Gamma_3 \Gamma_1^* \Gamma_2 + D_{\Gamma_3} \Gamma_4 D_{\Gamma_2} \\
\end{array}
\right],
\]

\noi then $\Gamma A$ corresponds to the following figure:

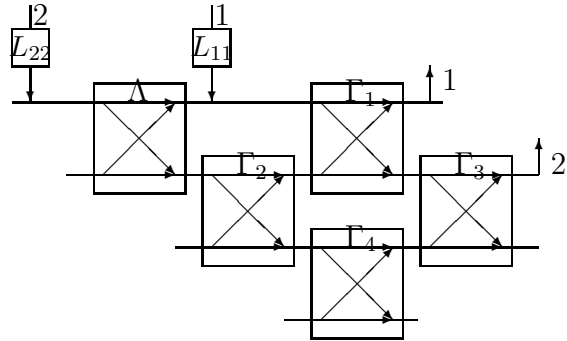
\begin{figure}[h]
\setlength{\unitlength}{3000sp}%
\begingroup\makeatletter\ifx\SetFigFont\undefined%

\gdef\SetFigFont#1#2#3#4#5{%
  \reset@font\fontsize{#1}{#2pt}%
  \fontfamily{#3}\fontseries{#4}\fontshape{#5}%
  \selectfont}%
\fi\endgroup%

\begin{picture}(6174,2949)(289,-2323)

{ \put(601, 439){\line(0, -1){200}}
}%

{ \put(451,-61){\framebox(300,300){\small $L_{22}$}}
}%

%input arrow 2
{ \put(601, -61){\vector(0, -1){300}}
}%
{ \put(621, 279){$2$}
}%

%upper line to the left of Box 1
{ \put(451,-361){\line( 1, 0){600}}
}%

%Box 1
{ \put(1401,-351){$\Lambda$}
}%

{ \put(1201,-361){\vector( 1,-1){600}}
}%
{ \put(1201,-961){\vector( 1, 1){600}}
}%
{ \put(1126,-1111){\framebox(750,900){}}
}%
{ \put(901,-961){\line( 1, 0){1200}}
}%
{ \put(901,-361){\line( 1, 0){1200}}
}%
{ \put(1201,-361){\vector( 1, 0){600}}
}%
{ \put(1201,-961){\vector( 1, 0){600}}
}%

{ \put(1951,-61){\framebox(300,300){\small $L_{11}$}}
}%
{ \put(2101, 439){\line(0, -1){200}}
}%
%input arrow 1
{ \put(2101, -61){\vector(0, -1){300}}
}%
{ \put(2121, 279){$1$}
}%

%upper line to the left of Box 2
{ \put(2001,-361){\line( 1, 0){1400}}
}%

%lower line connecting Box 1 and Box 2
{ \put(2101,-961){\line( 1, 0){1200}}
}%

% the vector on the above line
{ \put(2101,-961){\vector( 1, 0){600}}
}%

%Box2
{ \put(3201,-351){$\Gamma_1$}
}%
{ \put(2926,-1111){\framebox(750,900){}}
}%
{ \put(3001,-361){\vector( 1, 0){600}}
}%
{ \put(3001,-961){\vector( 1, 0){600}}
}%
{ \put(3001,-961){\vector( 1, 1){600}}
}%
{ \put(3001,-361){\vector( 1,-1){600}}
}%

%upper line connecting Box 2 and Box 3
{ \put(3601,-361){\line( 1, 0){400}}
}%

%output arrow 1
{ \put(3901,-361){\vector(0, 1){300}}
}%
{ \put(4001,-261){$1$}
}%

%lower line connecting Box 2 and Box 3
{ \put(3301,-961){\line( 1, 0){1200}}
}%

%vector on the above line
{ \put(3901,-961){\vector( 1, 0){600}}
}%

%upper line to the left of Box 4
{ \put(1801,-961){\line( 1, 0){3000}}
}%

%Box4
{ \put(2301,-951){$\Gamma_2$}
}%
{ \put(2026,-1711){\framebox(750,900){}}
}%
{ \put(2101,-1561){\vector( 1, 0){600}}
}%
{ \put(2101,-1561){\vector( 1, 1){600}}
}%
{ \put(2101,-961){\vector( 1,-1){600}}
}%

%lower line connecting Box 4 and Box 5
{ \put(1801,-1561){\line( 1, 0){3000}}
}%

%Box5
{ \put(4101,-951){$\Gamma_3$}
}%
{ \put(3826,-1711){\framebox(750,900){}}
}%
{ \put(3901,-961){\vector( 1,-1){600}}
}%
{ \put(3901,-1561){\vector( 1, 0){600}}
}%
{ \put(3901,-1561){\vector( 1, 1){600}}
}%

%output arrow 2
{ \put(4801,-961){\vector( 0, 1){300}}
}%
{ \put(4901,-951){$2$}
}%

%Box6
{ \put(3201,-1551){$\Gamma_4$}
}%
{ \put(2926,-2311){\framebox(750,900){}}
}%
{ \put(3001,-1561){\vector( 1, 0){600}}
}%
{ \put(3001,-2161){\vector( 1, 0){600}}
}%
{ \put(3001,-2161){\vector( 1, 1){600}}
}%
{ \put(3001,-1561){\vector( 1,-1){600}}
}%

%lower line
{ \put(2701,-2161){\line( 1, 0){1100}}
}%

\end{picture}

\caption{The $2 \times 2$ matrix $\Gamma A$}
\end{figure}

\section{Applications}

Due the the ubiquity of positive matrices, the Schur-Constantinescu parametrization of positive matrices has numerous applications \cite{Co}.
More recently, it has been applied in the context of quantum information theory. For example, it was used to parametrize
completely positive maps (quantum channels) in \cite{tiviII}. A cylinder-like condition, called the {\sl Bloch cylinder}, was obtained for 
positive matrices of trace 1 (quantum states) of any finite dimension. This provides an alternative to the well-known Bloch sphere. In \cite{Tseng2} 
it was applied to show that every positive map is completely positive to a certain extent, thus establishing the separability of certain families of 
quantum states in arbitrary finite dimensions. In a similar vein, in this section we obtain more results in this direction, in a sense extending what was 
found in \cite{Choi}. Also, we consider two further applications that are matrix completion problems in disguise and can be solved by utilizing parametrization of 
matrix contractions.

\subsection{Positive Maps}

In this section, the structure of contractions and positive matrices are applied to extend properties of positive maps.

\begin{defi} Let $\cH$ and $\cK$ be Hilbert spaces. A linear map $\Phi : \cL (\cH) \rightarrow \cL (\cK)$ is said to be {\bf positive} if it preserves the
cone of positive elements, i.e. $A \geq 0$ implies $\Phi(A) \geq 0$. Let $\bC ^{n \times n}$ denote the $n \times n$ matrices of complex numbers
and $I_n$ the identity map on $\bC ^{n \times n}$, then a map $\Phi$ is said to be {\bf n-positive} if the induced map

\[
I_d \otimes \Phi : \bC ^{n \times n} \otimes \cL (\cH) \rightarrow \bC ^{n \times n} \otimes \cL (\cK)
\]

\noi is positive, and $\Phi$ is {\bf completely positive}, or {\bf CP}, if it is $n$-positive for all $n$.
\end{defi}

We state the following result without proof \cite{Paulsen}.

\begin{thm}{\rm (Russo-Dye)} Let $\Phi$ be a positive map between unital C*-algebras, then $\| \Phi \| \leq \| \Phi(I) \|$.
\end{thm}
 
In particular, if $\Phi$ is unital and $\Gamma$ a contraction, then 

\[
\Phi (\Gamma^*) \Phi(\Gamma) = \| \Phi(\Gamma) \|^2 \leq \Gamma^* \Gamma \leq I.
\]

\noi Similarly, $\Phi (\Gamma) \Phi(\Gamma^*) = \leq I$. Making use of this, one has \cite{Choi}:

\begin{thm} Let $\Phi: \cL(\cH) \rightarrow \cL(\cK)$ be a postive map, then for all positive

\[
\rho =
\left[
\begin{array}{cc}
T & S \\
S^* & T
\end{array}
\right]
\in \bC ^{2 \times 2} \otimes \cL(\cH), 
\quad \mbox{we have} \quad (I_2 \otimes \Phi)(A) =  
\left[
\begin{array}{cc}
\Phi(T)   & \Phi(S) \\
\Phi(S^*) & \Phi(T)
\end{array}
\right]
\geq 0.
\]

\end{thm}

We recast the proof so that the role played by contractions is made more apparent.

\pf Assume for the moment that $T^{-1}$ exists and $\Phi(I)$ is invertible, therefore so is $\Phi(T)$. According to theorem 5, 
$S = T^{\frac{1}{2}} \Gamma T^{\frac{1}{2}}$ for some contraction $\Gamma$. It is equivalent to show that the Schur complement

\[
\Phi(T) - \Phi(S^*) \Phi(T)^{-1} \Phi(S) \geq 0 
\]

\noi i.e.

\[
I \geq \Phi(T)^{- \frac{1}{2}} \Phi(T^{\frac{1}{2}} \Gamma^* T^{\frac{1}{2}}) \Phi(T)^{- \frac{1}{2}} \cdot 
\Phi(T)^{- \frac{1}{2}} \Phi(T^{\frac{1}{2}} \Gamma T^{\frac{1}{2}}) \Phi(T)^{- \frac{1}{2}}.
\]

\noi This suggests that we define $\Psi :\cL(\cH) \rightarrow \cL(\cK)$ by

\[
\Psi(A) = \Phi(T)^{- \frac{1}{2}} \Phi(T^{\frac{1}{2}} A T^{\frac{1}{2}}) \Phi(T)^{- \frac{1}{2}}.
\]

\noi $\Psi$ is an unital positive map. By Russo-Dye,

\[
I \leq \Psi(\Gamma) \Psi (\Gamma^*) = \Phi(T)^{- \frac{1}{2}} \Phi(T^{\frac{1}{2}} \Gamma^* T^{\frac{1}{2}}) \Phi(T)^{- \frac{1}{2}} \cdot 
\Phi(T)^{- \frac{1}{2}} \Phi(T^{\frac{1}{2}} \Gamma T^{\frac{1}{2}}) \Phi(T)^{- \frac{1}{2}},
\]

\noi which is what we want.

If $T$ is not invertible, consider the sequence $\{ T_n = T + \frac{1}{n} \}$. $T_n$ tends to $T$ uniformly and positive maps are bounded. So

\[
\Phi(S) = \lim_n \Phi(T_n) ^{\frac{1}{2}} \cdot \Lambda_n \cdot \lim_n \Phi(T_n) ^{\frac{1}{2}}
\]

\noi for contractions ${\Lambda_n}$. Let $\Lambda \in \cL(\cK)$ be a weak operatorial limit of ${\Lambda_n}$, then

\[
\Phi(S) =  \Phi(\lim_n T_n) ^{\frac{1}{2}} \cdot \Lambda \cdot \Phi(\lim_n T_n) ^{\frac{1}{2}}
=  \Phi(T) ^{\frac{1}{2}} \Lambda \Phi(T) ^{\frac{1}{2}}.
\]

\noi So the claim holds.
 
If, in addition, $\Phi(I)$ is not invertible, take a positive functional $\phi$ with $\phi(I) = 1$. Define $\Phi_n (A) = \Phi(A) + \frac{1}{n} \phi(A)$. 
We have $\Phi_n \rightarrow \Phi$ in the operator norm of linear maps, and

\[
\Phi(S) = \lim_n \Phi_n (S) = \lim_n \Phi_n(T) ^{\frac{1}{2}} \cdot \Lambda'_n \cdot \lim_n \Phi_n(T) ^{\frac{1}{2}}.
\]

\noi The same weak limit argument shows that $\Phi(S) =  \Phi(T) ^{\frac{1}{2}} \Lambda' \Phi(T) ^{\frac{1}{2}}$ for some contraction $\Lambda'$. This
proves the theorem. $\Box$

In other words, any positive map is $2$-positive on the $2 \times 2$ Toeplitz matrices. Now we extend this to a sub-family of $3 \times 3$ positive 
matrices. Recall that an operator $A \in \cL(\cH)$ is said to be {\bf subnormal} if it is the compression of a $2 \times 2$ normal upper-triangular $N$, 
i.e. if there exist some Hilbert space $\cK$ and a normal $N \in \cL(\cK)$ such that $N$ is of the form
\[
N =
\left[
\begin{array}{cc}
A & B \\
0 & C \\
\end{array}
\right]. 
\]

The following fact, which we state without proof, will be used \cite{Choi} :

\begin{lem}
For any unital positive map $\Phi: \cL(\cH) \rightarrow \cL(\cK)$ and any normal $A \in \cL(\cH)$,

\[
\Phi(A^* A) \geq \Phi(A^*) \Phi(A) \quad \mbox{and} \quad \Phi(A^* A) \geq \Phi(A) \Phi(A^*).
\]
\end{lem}

What is known as {\sl Kadison's inequality} will also be needed: for every unital positive map $\Phi$ and every self-adjoint $S$,
$\Phi(S^2) \geq \Phi(S)^2$ \cite{Kadison}. What we will show that is essentially every positive map is $3-$positive in a certain limited sense. 
We first notice that subnormal contractions enjoy a property stronger than that prescribed by Russo-Dye.

\begin{lem}
If $\Phi$ is a unital positive map and $\Gamma$ a subnormal contraction, then

\[ 
I - \Phi(\Gamma^* \Gamma) - \Phi(D_{\Gamma^*})^2 \geq 0
\quad \mbox{and} \quad
I - \Phi(\Gamma^* \Gamma) - \Phi(D_{\Gamma})^2 \geq 0.
\]
\end{lem} 

\pf We directly compute

\[
I - \Phi(\Gamma^* \Gamma) - \Phi(D_{\Gamma^*})^2
\geq I - \Phi(\Gamma^* \Gamma) - \Phi(D_{\Gamma^*}^2)
\]
\[
= I - \Phi(\Gamma^* \Gamma) - \Phi(I) + \Phi( \Gamma ^* \Gamma)
= \Phi( \Gamma ^* \Gamma) - \Phi(\Gamma^* \Gamma),
\]

\noi which is positive, by the preceding lemma. The second inequality is similar. $\Box$

\begin{thm}
i) Consider positive matrices in $(A_{ij}) \in \bC^{3 \times 3} \otimes \cL(\cH)$ that are SC-parametrized in the following way: for $i = 1, 2, 3$,
choose $A_{ii} = T \geq 0$. Of the three contractions, choose $\Gamma_{12}$ to be subnormal, $\Gamma_{23} = 0$, and $\Gamma_{13} = I$. In other words, we 
consider $3 \times 3$ positive matrices of the form

\[
(A_{ij}) = 
\left[
\begin{array}{ccc}
T        & T^{\frac{1}{2}} \Gamma T^{\frac{1}{2}} & T^{\frac{1}{2}}  D_{\Gamma^*} T^{\frac{1}{2}} \\
T^{\frac{1}{2}} \Gamma^* T^{\frac{1}{2}} & T & 0\\
T^{\frac{1}{2}} D_{\Gamma^*} T^{\frac{1}{2}} & 0 & T
\end{array}
\right].
\] 

\noi Then

\[
(I_3 \otimes \Phi)(A_{ij}) \geq 0,
\]

\noi for any positive map $\Phi$ acting on $\cL(\cH)$.

ii) The same is true if $\Gamma_{23}$ is subnormal, $\Gamma_{12} = 0$, $\Gamma_{13} = I$, i.e. if
  
\[
(A_{ij}) = 
\left[
\begin{array}{ccc}
T                                             & 0  & T^{\frac{1}{2}}  D_{\Gamma} T^{\frac{1}{2}} \\
0                                             & T  & T^{\frac{1}{2}} \Gamma T^{\frac{1}{2}}\\
T^{\frac{1}{2}} D_{\Gamma} T^{\frac{1}{2}}    & T^{\frac{1}{2}} \Gamma^* T^{\frac{1}{2}} & T
\end{array}
\right].
\] 
\end{thm}

We do not completely recover Choi's result by considering the $2 \times 2$ leading minor in part i). The requirement that $\Gamma$ be subnormal is 
particular to the $3 \times 3$ case, due to the presence of $D_{\Gamma^*}$ in the parametrization.\\

\pf i) Assume for the moment that $\Phi(I)$ and $T$ are invertible. Invoking again theorem 5 on the $2 \times 2$ case, it is equivalent to show that

\[
\Phi(T) \geq
\left[
\begin{array}{cc} \Phi(T^{\frac{1}{2}} \Gamma T^{\frac{1}{2}}) & \Phi(T^{\frac{1}{2}}  D_{\Gamma^*} T^{\frac{1}{2}})\end{array}
\right]
\left[
\begin{array}{cc}
\Phi(T)^{-1} & 0\\
0 & \Phi(T)^{-1}
\end{array}
\right]
\left[
\begin{array}{c}
\Phi(T^{\frac{1}{2}} \Gamma^* T^{\frac{1}{2}}) \\
\Phi(T^{\frac{1}{2}} D_{\Gamma^*} T^{\frac{1}{2}})
\end{array}
\right].
\]

\noi The right hand side is

\[
\Phi(T^{\frac{1}{2}} \Gamma T^{\frac{1}{2}}) \Phi(T)^{-1} \Phi(T^{\frac{1}{2}} \Gamma^* T^{\frac{1}{2}}) +
\Phi(T^{\frac{1}{2}} D_{\Gamma^*} T^{\frac{1}{2}}) \Phi(T)^{-1} \Phi(T^{\frac{1}{2}} D_{\Gamma^*} T^{\frac{1}{2}}).
\]

\noi Again we define an unital positive map $\Psi$ by

\[
\Psi(B) = \Phi(T)^{-\frac{1}{2}} \Phi(T^{\frac{1}{2}} B T^{\frac{1}{2}}) \Phi(T)^{-\frac{1}{2}}.
\]
 
\noi By the subnormality of $\Gamma$ and lemma 4, 

\[
I \geq \Psi(\Gamma) \Psi(\Gamma^*) + \Psi(D_{\Gamma^*})^2
\]

\noi whicn proves the claim. The general case can be shown using argument similar in that of theorem 7.  

ii) The argument is analogous to i) and omitted. $\Box$\\

The following result is of similar nature. The special $2 \times 2$ case, proven in \cite{Choi}, says that any positive map is positive on $2 \times 2$ 
Hankel matrices.

\begin{thm} If $A$ is a positive matrix of the form

\[
A =
\left[
\begin{array}{ccccc}
T      & 0      & \cdots & 0      & S_1 \\
0      & T      & \cdots & 0      & S_2 \\
\vdots & \vdots & \ddots & \vdots & \vdots\\
0      & 0      & \cdots & T      & \vdots\\
S_1    & S_2    & \cdots & 0      &R
\end{array}
\right]
\quad \mbox{or} \quad
\left[
\begin{array}{ccccc}
T      & S_1    & \cdots & \cdots & S_{m-1}\\
S_1    & R      & 0      & \cdots & 0 \\
\vdots & 0      & \ddots & \cdots & \vdots \\
\vdots & \vdots & \vdots & R      & 0\\
S_{m-1}& 0      & \cdots & 0      & R
\end{array}
\right]
\in C^{m \times m} \otimes \cL(\cH),
\] 

\noi where $m$ is arbitrary, then 

\[
(I_m \otimes \Phi) (A) \geq 0
\] 

\noi for any positive map $\Phi$. 
\end{thm}

\pf Suppose $A$ is of the first form in the claim. The argument uses only the structure of $2 \times 2$ matrices and thus we consider first the 
case $m = 2$. By virtue of previous arguments, it can be assumed without loss of generality that $T$ and $\Phi(I)$ are both invertible. To show

\[
\left[
\begin{array}{cc}
\Phi(T) & \Phi(S) \\
\Phi(S) & \Phi(R) \\
\end{array}
\right]
\geq 0
\]

\noi, it suffices to obtain $\Phi(S T^{-1} S) \geq \Phi(S) \Phi(T)^{-1} \Phi(S)$ because the Schur complement $R -S T^{-1} S$ is positive. This
is equivalent to 

\[
\Phi(T) ^{-\frac{1}{2}} \Phi(S T^{-1} S) \Phi(T) ^{-\frac{1}{2}} \geq
\Phi(T) ^{-\frac{1}{2}} \Phi(S) \Phi(T) ^{-\frac{1}{2}} \cdot \Phi(T) ^{-\frac{1}{2}} \Phi(S) \Phi(T) ^{-\frac{1}{2}},
\]

\noi
which suggests we define a unital positive map by

\[
\Psi(B) = \Phi(T) ^{-\frac{1}{2}} \Phi(B) \Phi(T) ^{-\frac{1}{2}}.
\]

\noi
Invoking Kadison's inequality then proves the $2 \times 2$ case. For $m > 2$, it is enough to show that

\[
\Phi(
\left[
\begin{array}{ccc} S_1 & \cdots & S_{m-1} \end{array}
\right]
\left[
\begin{array}{ccc} 
T^{-1} & \cdots & 0\\
\vdots & \ddots & \vdots\\
0      & \cdots & T^{-1}
\end{array}
\right]
\left[
\begin{array}{c} S_1 \\ \cdots \\ S_{m-1} \end{array}
\right])
\]

\[
\geq \sum_{i = 1} ^{m-1} \Phi(S_i) \Phi(T)^{-1} \Phi(S_i),
\]

\noi i.e.

\[
\Phi(\sum_{i = 1} ^{m-1} S_i T^{-1} S_i) \geq \sum_{i = 1} ^{m-1} \Phi(S_i) \Phi(T)^{-1} \Phi(S_i).
\]

\noi But $A$ is positive only if the principal $2 \times 2$ minors 

\[
\left[
\begin{array}{cc}
T & S_i \\
S_i & R \\
\end{array}
\right]
\]

\noi are positive. Thus the desired inequality holds by linearity of $\Phi$ and the $2 \times 2$ argument. $\Box$\\

\begin{rmk} What we have show above is that a positive map is $3$-positive and CP (in the case of theorems 8 and 9 respectively) to
a certain extent. Results in the similar vein were obtained in \cite{Tseng2} that are also applications of SC parameters. Namely positive maps
were shown to be CP if restricted to certain families, of arbitrary finite size, which can be SC-parametrized by two real parameters. In that approach
, Choi's result on $2 \times 2$ Hankel matrices also become a special case. For comparison, a result from \cite{Tseng2} for matrices with entries in 
$\bC^{3 \times 3}$ is stated below.
\end{rmk}

\begin{thm} Let $\cS$ be the linear span of
\[
\{
A =
\left[
\begin{array}{ccc}
a & a & b\\
a & a & b\\
b & b & c
\end{array}
\right]
| a, b, b \in \bC \} \subset \bC^{3 \times 3}
\]

\noi and where $m \in \bN$ be arbitary. Then for all positive $\rho \in \bC^{m \times m} \otimes \cS$, 

\[
(I_m \otimes \Phi)(\rho) \geq 0
\]

\noi where $\Phi$ is any positive map acting on $\bC^{3 \times 3}$. Furthermore, the claim holds if $A$ is replaced by

\[
\left[
\begin{array}{ccc}
a & c & a\\
c & b & c\\
a & c & a
\end{array}
\right]
\quad \mbox{or} \quad
\left[
\begin{array}{ccc}
a & c & c\\
c & b & b\\
c & b & b
\end{array}
\right].
\]
\end{thm}

\noi {\em Separable Quantum States} In physical language, trace-class positive matrices are un-normalized mixed states 
\cite{Werner}. 

\begin{defi}\cite{Werner2}
Let the state space of a bipartite quantum system be the tensor product $\cH = \cH_1 \otimes \cH_2$, where $\cH_i$ are Hilbert spaces.
A state $\sigma \in \cL(\cH)$ is {\bf separable} if it lies in the closure, in the trace norm, of states of the form

\[
\rho = \sum_{i = 1} ^k \rho^1_i \otimes \rho^2_i ,
\] 
 
\noi where $\rho^j _i$ are states in $\cH_j$.
\end{defi}

The membership problem for separable states is sometimes called the {\sl separability problem}.
The following theorem establishes the correspondence between the classification of positive maps and the 
membership problem for separable states\cite{Horodecki}:\\

\begin{thm}If a mixed state $\sigma \in {\mathcal L}( {\mathcal H}_A ) 
\otimes {\mathcal L}( {\mathcal H}_B )$ is such that
for every positive map $\Phi$ from ${\mathcal L}({\mathcal H}_B)$ to 
${\mathcal L}({\mathcal H}_A)$, the operator $(I_A \otimes \Phi)(\sigma)$ 
is positive, then $\sigma$ is separable.\end{thm}

The above result is of geometric nature and a consquence of the hyperplane-separation variant of Hahn-Banach.
Thus in quantum information theory, positive but not CP maps are called {\sl entanglement witnesses}, for they detect the entanglement
of some state. If a family $\cS$ of positive matrices is such that any positive map behaves as a CP map when restricted to $\cS$, then
$\cS$ must consist of separable states, due to lack to entanglement witnesses. Thus what was shown in the previous discussion translate to
that all positive matrices of the forms specified in theorems 8 and 9 are separable states. In particular, $2 \times m$ block-Toeplitz and block-Hankel 
states are separable, where $m$ need not be finite.

\subsection{POVM's}

We first give a few relevant definitions and basic results; the reader is referred to \cite{Werner} for more background information. 
In the von Neumann measurement scheme, the {\sl effects} of a quantum measurement are assumed to satisfiy the {\sl projective hypothesis}, 
i.e. they are self-adjoint projections and form the resolution of identity of a self adjoint operator. A resolution of the identity is sometimes called 
{\sl projection-valued measure}, or {\sl PVM}. A more general formulation of measurement replaces these projections by 
positive operators:

\begin{defi}\cite{Paulsen} Let $X$ be a compact Hausdorff space, $\cB$ be the Borel $\sigma-$algebra on $X$, and $\cH$ a Hilbert space. 
A {\bf positive operator-valued measure}, or {\bf POVM} is a map $E: \cB \rightarrow \cL(\cH)$ such that $E(B) \geq 0$ for all $B$, and 
$E$ is countably additive in the weak topology on $\cH$, i.e. for any pairwise disjoint collection $\{B_i\}_{i \geq 1} \subset \cB$,

\[
\la E(\cup_i B_i) x, y \ra = \sum_i \la E(B_i) x, y \ra 
\]

\noi for all $x, y \in \cH$.
\end{defi}

If $E(B)$ is self adjoint for all $B$ and $E(B_1 \cap B_2) = E(B_1) E(B_2)$, then each $E(B)$ is a self adjoint projection. Thus $E$ is a PVM and we recover
the von Neumann Scheme. A natural question is whether a POVM can be "lifted" to a larger space where it is a PVM. The general answer to this 
dilation-theoretic question is given by:

\begin{thm}{\rm (Naimark)} Let $X$ be a compact Hausdorff space. Suppose $E$ is a POVM on the $\sigma$-algebra generated 
by the Borel sets of $X$ and $E$ takes values in $\cL (\cH)$. There exist a Hilbert space $\cK$ containing ${\cH}$ 
as a subspace and a PVM $F$ on $X$ with values in $\cL (\cK)$ such that 

\[
E(B) = P_{\cH} F(B) P_{\cH}
\]

\noindent ,for all Borel $B \subset X$.
\end{thm}

The proof follows from the fact that $C(X)$ is a commutative C*-algebra and therefore the induced map

\[
f \in C(X) \stackrel{\Phi}{\longrightarrow} \int _X f(x) d E(x)
\]

\noindent is completely positive, rather than merely positive. Stinespring's theorem on CP maps \cite{Stinespring} then says $\Phi$ can be dilated to a
homomorphism $\Phi '$. The PVM corresponding to $\Phi '$ is the desired $F$. For a complete proof, see \cite{Paulsen}.
Stinespring's theorem is a generalization of the Gelfand-Naimark representation thereom of postive functionals.\\
 
%For instance, if one is to view Choi's theorem on completely positive maps as a direct corollary of Stinespring's result, the explicit form of the Kraus operators would not be obtained directly. 

In quantum information theory, Of particular interest is the case when $X$ is finite, with the discrete topology. In that case, one would like and can 
indeed find solutions of more concrete nature. Let $X = \{1, 2, \cdots , n\}$. Without losing generality, we can consider only POVM's whose elements 
are rank-1 projections that may not be mutually orthogonal. Suppose a POVM on $X$ is given by $E(i)= v_i v_i ^*$, $i = 1, \cdots, n$ with 
$\sum _i v_i v_i ^* = I_m$ where $m \leq n$ and $I_m$ is the identity in $\bC ^{m \times m}$. In other words, 

\[
M = \left[ \begin{array}{ccc}v_1 & \cdots & v_n \end{array} \right] \in \bC ^{m \times n}
\]

\noindent is an isometry, i.e. $M^*M = I$. We want to specify a PVM $F$ taking value in some $\cL( \cK)$ whose restriction to $\bC ^m$ is $E$. 
%In Naimark's theorem, the C*-algebra $C(X)$ is fixed. 
This is a trivial completion problem: given a (rectangular) isometry $M$, find a 
suitable rectangular $N$ s.t. $\left[ \begin{array}{c} M \\ N \end{array} \right]$ is unitary. It is an elementary fact from linear algebra that such 
an $N$ can always be found.\\

However, in physical considerations, a suitable POVM is often obtained by coupling to the original system an {\sl ancilla}. This amounts to
finding appropriate operators $A$, $B$, and $C$ such that 

\[
U=
\left[
\begin{array}{cc}
M & A \\ B & C 
\end{array}
\right]
\in \bC ^{k \times k}
\]

\noindent is unitary. The columns of U, $\{ u_1, \cdots, u_k \}$, then gives a PVM on $\bC^k = \bC^m \oplus \bC^{k-m}$ with the desired 
properties.\\

The completion of $2 \times 2$ unitary matrices offers an immediate solution to this problem. Since $M$ is an isometry, 
it is trivially a contraction. For example, consider the special case of Julia operator. So the PVM are the projections onto the column vectors of 

\[
J(M)=
\left[
\begin{array}{cc}
M   & D_{M^*} \\ 
D_M & -M^* 
\end{array}
\right]
\] 

\noindent
$M^*M = I$ means $D_M = 0$ and similarly $D_{M^*} = (I - M M^*)^{\frac{1}{2}} = (I - M M^*)$ because $I - M M^*$ is a projection. 
Thus 

\[
J(M)=
\left[
\begin{array}{cc}
M   & I - M M^* \\ 
0   & -M^* 
\end{array}
\right]
\] 

\noindent and a suitable PVM $F$ is obtained without doing any calculation whatsoever.  If $M$ is $m \times n$ with $m \leq n$, the 
Hilbert space $\cK$ is of dimension $m+n$. From the discussion on unitary operators, we see that the freedom in obtained a suitable PVM is 
described by

\[
J'(M)=
\left[
\begin{array}{cc}
I   & \\ 
0   & U_1 
\end{array}
\right]
\left[
\begin{array}{cc}
M   & I - M M^* \\ 
0   & -M^* 
\end{array}
\right]
\left[
\begin{array}{cc}
I   & \\ 
0   & U_2 
\end{array}
\right]
\]

\noi where $U_i$ are unitary matrices of suitable size. An effect of the POVM can be obtained
by first performing the corresponding yes-no measurement from the dilated PVM then taking the partial trace with respect to the ancilla 
variables.\\

\noi {\em Note} Physical reasons require that the dimension of $J'(M)$ be $(m \times k) \times (m \times k)$ where the ancilla has state space of 
dimension $k$. This can always be achieved via direct sum with an identity matrix of appropriate size.

\subsection{Mocking Up a Quantum Operation}

We now extend our discussion to general quantum operations, of which measurement is a special case. We are interested in the finite dimensional case.

\begin{defi}
A quantum operation is a completely positive map $\Phi : \bC ^{n \times n} \rightarrow \bC ^{m \times m}$ between density matrices that does not increase 
the trace.  
\end{defi}

It is well known that $\Phi$ must take the form $\Phi (\rho) = \sum _{i = 1} ^{nm} E_i \rho E_i ^*$ \cite{Choi1}. 
By {\sl mocking up} \cite{nie} we mean, similar to the POVM case, coupling the system to an ancilla and find an unitary evolution on the combined system such that 
the reduced state, obtained via the partial trace, is $\sum _{i = 1} ^{nm} E_i \rho E_i ^*$. In other words, given a CP map $\Phi$, we wish
to find an unitary dilation. Again Stinespring's theorem ensures the existence of such a dilation. But in practice perhaps one would like to obtain a less
abstract solution. The parametrization of matrix contrations provides one such explicit and easy procedure.\\

Choose the ancilla to be $\bC ^{nm}$. Any quantum state can be purified \cite{nie}, and we can assume the ancilla is in a pure state, a rank-1 
projection, of the special form $e_0 e_0 ^*$, where

\[
e_0 = \left[ \begin{array}{c} 1 \\ 0 \\ \vdots \\ 0 \end{array} \right].
\]
 
\noindent
The state of the composite system is

\[
e_0 e_0 ^* \otimes \rho = 
\left[ 
\begin{array}{ccc} 
\rho &  0  & \; \\ 
0    &  0  & \; \\
\;   &  \; & \ddots
\end{array}\right].
\] 

\noindent
If $U$ is the proposed unitary evolution, then

\[
U \rho \; U^* =
\left[ 
\begin{array}{ccc} 
U_{11} &  U_{12}  & \; \\ 
U_{21} &  U_{22}  & \; \\
\;   &  \; & \ddots
\end{array}\right]
\left[ 
\begin{array}{ccc} 
\rho &  0  & \; \\ 
0    &  0  & \; \\
\;   &  \; & \ddots
\end{array}\right]
\left[ 
\begin{array}{ccc} 
U_{11}^*   &  U_{21}^*  & \; \\ 
U_{12}^*   &  U_{22}^*  & \; \\
\;         &  \;        & \ddots
\end{array}\right]
\]

\[
=
\left[ 
\begin{array}{ccc} 
U_{11} \rho U_{11}^* &  U_{11} \rho U_{21}^*  & \; \\ 
U_{21} \rho U_{11}^* &  U_{21} \rho U_{21}^*  & \; \\
\;   &  \; & \ddots
\end{array}\right].
\]

\noindent Tracing out the ancilla, the first system, gives the reduced density matrix $\sum _1 ^{nm}  U_{i1} \rho U_{i1}^*.$ Therefore
to specify $U$ is to find appropriate operators $A$, $B$, and $C$ such that 

\[
U=
\left[
\begin{array}{cc}
T & A \\ B & C 
\end{array}
\right]
\in \bC^{k \times k}
\]

\noindent is unitary, where $T$ is the contraction

\[
T = \left[ \begin{array}{c} E_1 \\ \vdots \\ E_{nm} \end{array} \right].
\]

\noindent As before, this can be achieved by forming the operator \\

\[
J'(T)=
\left[
\begin{array}{cc}
I   & \\ 
0   & U_1 
\end{array}
\right]
\left[
\begin{array}{cc}
T   & D_{T^*} \\ 
D_T   & -T^* 
\end{array}
\right]
\left[
\begin{array}{cc}
I   & \\ 
0   & U_2 
\end{array}
\right].
\]

\noindent Notice purfication of mixed states was applied only to the ancilla. There is another approach that relies on purification 
more heavily. Namely, one treats the matrix $(\Phi(E_{ij}))_{ij}$, where $E_{ij}$ are standard matrix units, as a state and consider its purification. 
The differences are, first, in the latter only the range of $\Phi$ is coupled to an ancilla and, second, the lifted map still may not be an unitary 
evolution.


\begin{thebibliography}{99}
\frenchspacing

\bibitem{Werner} Alber,~G., et al., {\it Quantum Information},
Springer-Verlag, 2001.

\bibitem{Gheon} Arsene,~G.; Gheondea,~A., Completing matrix contractions,
{\em J. Operator Theory} {\bf 7},  179-189 (1982).
 

\bibitem{Choi1}
Choi,~M.~D., Completely positive linear maps on complex matrices, 
{\em Lin. Alg. Appl.} {\bf 10} 285-290 (1975).

\bibitem{Choi}
Choi,~M.~D., Some assorted inequalities for positive linear maps on C*-algebras,
{\em J. Operator Theory} {\bf 4}, 271-285 (1980).


\bibitem{Co}
Constantinescu,~T.,
{\it Schur Parameters, Factorization and Dilation Problems}, Birkh\"auser, 1996.

\bibitem{tiviII} Constantinescu,~T.; Ramakrishna,~V., Parametrizing quantum states and channels,
{\em Quantum Information Processing} {\bf 2}(3), 221-248 (2003).

\bibitem{Horodecki} Horodecki,~M.; Horodecki,~P.; Horodecki, ~R., Separability of mixed states: necessary and sufficient conditions,
{\em Phys. Lett. A} {\bf 223}, 1-8 (1996). 


\bibitem{Kadison} Kadison,~R.~V., A generalized Schwarz inequality and algebraic invariants for operator algebras,
{\em Ann. of Math.} {\bf 56} 494-503 (1952).

\bibitem{nie} Nielsen,~M.; Chuang,~I., {\it Quantum Computation and Quantum Information}, Cambridge University Press, 1999.


\bibitem{Paulsen}
Paulsen,~V., {\it Completely Bounded Maps and Operator Algebras},
Cambridge University Press, 2003.


\bibitem{Stinespring}
Stinespring,~W.~F., Positive maps on C*-algebras,
{\em Proc. Amer. Math. Soc.} {\bf 6}, 211-216 (1955).


\bibitem{Tseng2} Tseng,~M.~C.;Ramakrishna,~V., Dilation theoretic parametrizations of positive matrices with applications to quantum information,
preprint, quant-ph/0610021.

\bibitem{Werner2} Werner,~R.~F., Quantum states with Einstein-Podolsky-Rosen correlations admitting a hidden-variable Model, 
{\em Phys. Rev. A} {\bf 40}, 4277-4281 (1989).

  
\end{thebibliography}
\end{document}